\documentclass[12pt,a4paper]{article}
\usepackage{graphicx}
\usepackage{times}
\title{\bf High Mass X-ray Binaries in the NIR: Orbital solutions of two highly obscured systems. \footnote{Based on observations carried out at the European Southern Observatory under programmes 081.D-0073(A), 077.B-0872(A) and 081.D-0073(B).}}
\author{A.B. Mason$^1$\thanks{a.mason@open.ac.uk},A.J. Norton$^1$,J.S. Clark$^1$,P. Roche$^1$,I. Negueruela$^2$\\
\vspace{1cm}\\
\normalsize $^1$ The Open University, Milton Keynes, UK\\ 
\normalsize $^2$ Universidad de Alicante, Alicante, Spain}
\date{\mbox{}}
\begin{document}
\maketitle
\pagestyle{empty}
%
%
\def\bull{\vrule height .9ex width .8ex depth -.1ex}
\makeatletter
\def\ps@plain{\let\@mkboth\gobbletwo
\def\@oddhead{}\def\@oddfoot{\hfil\tiny\bull\quad
``The multi-wavelength view of hot, massive stars''; 39$^{\rm th}$ Li\`ege Int.\ Astroph.\ Coll., 12-16 July 2010 \quad\bull}%
\def\@evenhead{}\let\@evenfoot\@oddfoot}
\makeatother
%
%
\def\beginrefer{\section*{References}%
\begin{quotation}\mbox{}\par}
\def\refer#1\par{{\setlength{\parindent}{-\leftmargin}\indent#1\par}}
\def\endrefer{\end{quotation}}
%
%
{\noindent\small{\bf Abstract:} 
The maximum mass of a neutron star (NS) is poorly defined. Theoretical attempts to define this mass have thus far been unsuccessful. Observational results currently provide the only means of narrowing this mass range down. Eclipsing X-ray binary (XRB) pulsar systems are the only interacting binaries in which the mass of the NS may be measured directly. Only 10 such systems are known to exist, 6 of which have yielded NS masses in the range 1.06 - 1.86 M$_{\odot}$.We present the first orbital solutions of two further eclipsing systems, OAO 1657-415 and EXO 1722-363, whose donor stars have only recently been identified. Using observations obtained using the VLT/ISAAC NIR spectrograph, our initial work was concerned with providing an accurate spectral classification of the two counterpart stars, leading to a consistent explanation of the mechanism for spin period evolution of OAO 1657-415. Calculating radial velocities allowed orbital solutions for both systems to be computed. These are the first accurate determinations of the NS and counterpart masses in XRB pulsar systems to be made employing NIR spectroscopy.

}
%
%
\section{Introduction}
Despite extensive and ongoing theoretical work on the NS equation of state (EOS), the precise nature of the fundamental physical properties of NS matter is still poorly defined. Observational work can assist in reducing the number of contending theories by eliminating those that place unrealistic constraints on the mass range of observed NS. NS masses may only be determined from binary systems, within this paper we consider a specific class of these objects, those containing an eclipsing X-ray binary pulsar. At present only 10 such systems are known within our Galaxy. Prior to this work 6 of the NS in these systems have had mass determinations, with the donor star in each observable optically. Within this paper we discuss the first mass estimates found for NS employing near-infrared (NIR) spectroscopy. We have studied two eclipsing X-ray pulsar systems containing a High Mass donor, OAO 1657-415 and EXO 1722-363. Initially an accurate spectral classification for each of the donor stars was conducted utilising observations made using the VLT/ISAAC NIR spectrograph and current NIR spectral atlases. Using multi-epoch NIR spectra of each system we were able to determine the radial velocity of the donor star in each of the two systems thus enabling the construction of an orbital solution. This solution was then employed in calculating the mass estimate of each NS and the corresponding high mass donor, placing constraints upon the system inclination and separation of the binary system.
  
\section{Spectral classification} 

\subsection{Spectral classification of EXO 1722-363}
EXO 1722-363 (alternatively designated IGR J17252-3616) was discovered in 1984 by {\it EXOSAT} Galactic plane observations (Warwick et al. 1988).  {\it XMM-Newton} observations narrowed down the source position location to 4$^{\prime\prime}$. This allowed the identification of an IR counterpart 2MASS J17251139-3616575 (with magnitude J = 14.2, H = 11.8 and K$_{s}$ = 10.7 (Zurita Heras et al. 2006)). 
Examining Fig.1 we can see that all of  absorption lines in this spectrum are narrow, indicative of the object being a supergiant. EXO 1722-363 shows the singlet He\,{\sc i} 2.058 $\mu$m line in emission, this line being highly sensitive to wind and temperature properties. The N {\sc iii} 2.115 $\mu$m emission line is a common feature in B0-B1 supergiants. The absence of strong Br$\gamma$ 2.1655 $\mu$m emission features implies that EXO 1722-363 does not exhibit a strong stellar wind. 
From a qualitative comparison of spectra from Hanson et al. 2005, we identify EXO 1722-363 as being of spectral type B0-B1 Ia (Mason et al. 2009). 
By comparison with evolutionary rotational massive star models (Meynet \& Maeder, 2000) we find an initial progenitor mass for EXO 1722-363 in the range 30M$_{\odot}$ - 40M$_{\odot}$. Following the method for determining spectroscopic distance as detailed in Bibby et al. 2008, we determined a distance for EXO 1722-363 of $8.0_{-2.0}^{+2.5}$ kpc which is comparable within errors to the distance deduced in Thompson et al, 2007.
 Comparing our calculated distance with model fluxes derived from spectral fits to EXO 1722-363 (Corbet et al, 2005), we found that EXO 1722-363 has an intrinsic X-ray flux variability (in the range 2-60 keV) such that $F_{\rm min}$ = 0.78 ~ $\times$ 10$^{-10}$ erg cm$^{-2}$ s$^{-1}$ and $F_{\rm max}$ = 12.2 ~ $\times$ 10$^{-10}$ erg cm$^{-2}$ s$^{-1}$. We derive X-ray luminosities for EXO 1722-363 such that L$_{\rm {X_{min}}}$ = 3.4 $\times$ 10$^{35}$ erg s$^{-1}$ and L$_{\rm {X_{max}}}$ = 1.6 $\times$ 10$^{37}$ erg s$^{-1}$. We find this luminosity range entirely consistent with EXO 1722-363 being the donor within an SGXRB system.

\begin{figure}[h]
\centering
\includegraphics[width=12cm]{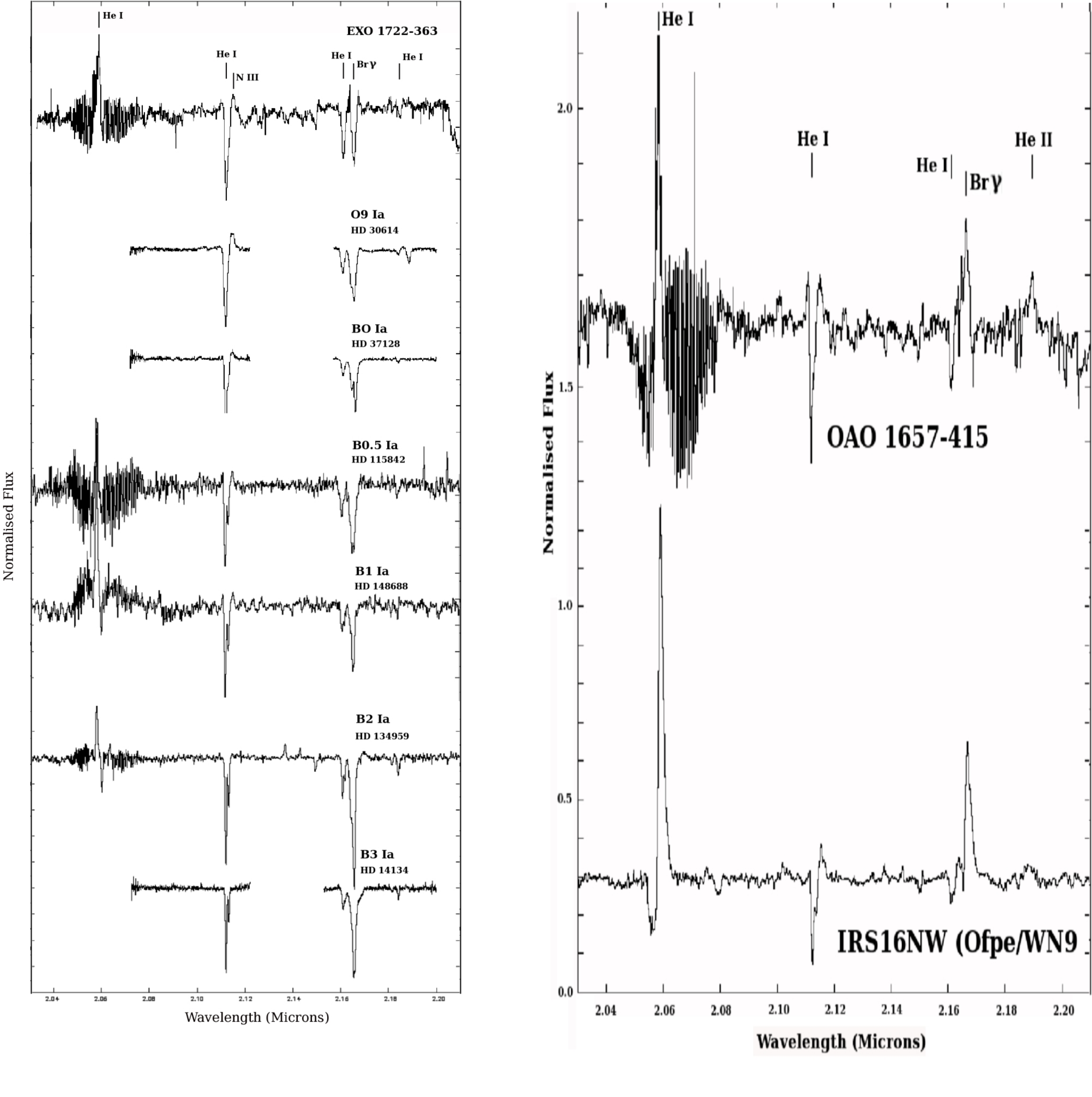}
\caption{Left Figure : Comparison of EXO 1722-363 and template O9-B3 Ia spectra from Hanson et al., 2005. Right Figure: Topmost shows a spectrum of OAO1657-415 compared with a Ofpe/WNL spectrum of IRS16NW from Martins et al., 2007.  }
\end{figure}

\subsection{Spectral classification of OAO 1657-415}

OAO 1657-415 was first detected over 30 years ago by the {\it Copernicus} X-ray satellite. 
From an examination of the orbital parameters of the X-ray pulsar it was determined that OAO 1657-415 was a high-mass system, indicating the mass of the donor lay between 14-18 M$_{\odot}$ with a corresponding radius range of 25-32R$_{\odot}$. Determination of these stellar parameters led to a suggested classification of B0-6 I (Chakrabarty et al,\ 1993).The correct identification of the donor in this system required the precise location of OAO 1657-415 to be accurately made. This was achieved by the {\it Chandra X-Ray Observatory} narrowing the X-ray location error radius down to 0.5$^{\prime\prime}$. Optical imaging of this position did not detect any donor candidates down to a magnitude of V$>$23. Near infrared imaging was employed to overcome significant levels of interstellar reddening, resulting in the identification of a donor located within the {\it Chandra} error radius. A corresponding IR counterpart was located in the {\it 2MASS} catalogue, 2MASS J17004888-4139214 (with magnitudes J = 14.1, H = 11.7 and K$_{\rm s}$ = 10.4) with A$_{V}$ = 20.4 $\pm$ 1.3, located at a distance of 6.4 $\pm$ 1.5 kpc (Chakrabarty et al,\ 2002). 
NIR K$_{\rm s}$ band spectroscopy of the donor obtained in 2008 (Mason et al,\ 2009) led to a re-evaluation of the spectral classification. Close examination revealed that OAO 1657-415 shared a similar spectral morphology with that of Ofpe/WNL stars. These are stars in transition between the OB main sequence and hydrogen depleted Wolf-Rayet stars, whose evolution follows from a wide range of progenitor masses.   
The spectrum of the mass donor in OAO1657-415 is presented in Fig. 1, and is dominated by He\,{\sc i} 
2.058 $\mu$m and Br$\gamma$ emission, the former stronger than the latter. We find a poor 
correspondence with the spectra of B0-6 supergiants (Hanson et al, 1996, Hanson et al, 2005) - as suggested for the 
mass donor by Chakrabarty et al,\ 2002 on the basis of a combination of photometric and X-ray data. However, comparison with the spectra of massive transitional objects presented by Morris et al,\ 1996 is more encouraging. In particular OAO 1657-415 shows pronounced similarities to the hot Ofpe/WNL 
stars. Consequently we may not {\em a priori} determine a unique distance to OAO 1657-415 on the basis of this classification. We thus find inevitably unconstructive limits of 4.4~kpc $<$ {\it d} $<$ 12~kpc. In turn this results in  1.5 $\times$ $10^{36}$ ~erg$ ^{-1} < ~ L_{\rm X} < 10^{37}$~erg s$^{-1}$, also entirely consistent with  observed luminosities of SGXRBs. Adopting the distance derived by Audley et al, 2006 leads to log(L/L$_{\odot}$) $\sim$ 5.7. For such a luminosity, comparison to the evolutionary tracks for massive stars (Meynet \& Maeder, 2000) 
imply an initial  mass of  $\sim$40~M$_{\odot}$.

\section {OAO 1657-415 : A mechanism for spin-period evolution}

We now turn to the implications of the Ofpe/WNL classification for the  X-ray properties of OAO 1657-415. The
 anomalous position of OAO 1657-415 within the Corbet diagram (Figure 2) (Corbet et al, 1986), is then naturally 
explained in terms of the properties of its  stellar wind. Compared to normal OB supergiants  
(Crowther et al,\ 2006), Ofpe/WNL stars typically demonstrate systematically  lower wind velocities and higher mass loss rates (Martins et al.\ 2007).
 This combination of wind properties permits a higher accretion rate and hence  transfer of angular momentum to the 
NS, in turn leading to a smaller (instantaneous) equilibrium spin period with respect to normal OB 
supergiants ($P_{\rm spin} 
\propto$ \.{M}$^{-3/7}$$v_{\infty}^{12/7}$ from Eqn. 12 of Waters et al, 1989,  where $P_{\rm spin}$, \.{M} 
and $v_{\infty}$ are the spin period of  the NS and the  mass loss rate and terminal velocity of the mass donor
 wind respectively).

\begin{figure}[h]
\centering
\includegraphics[width=7cm]{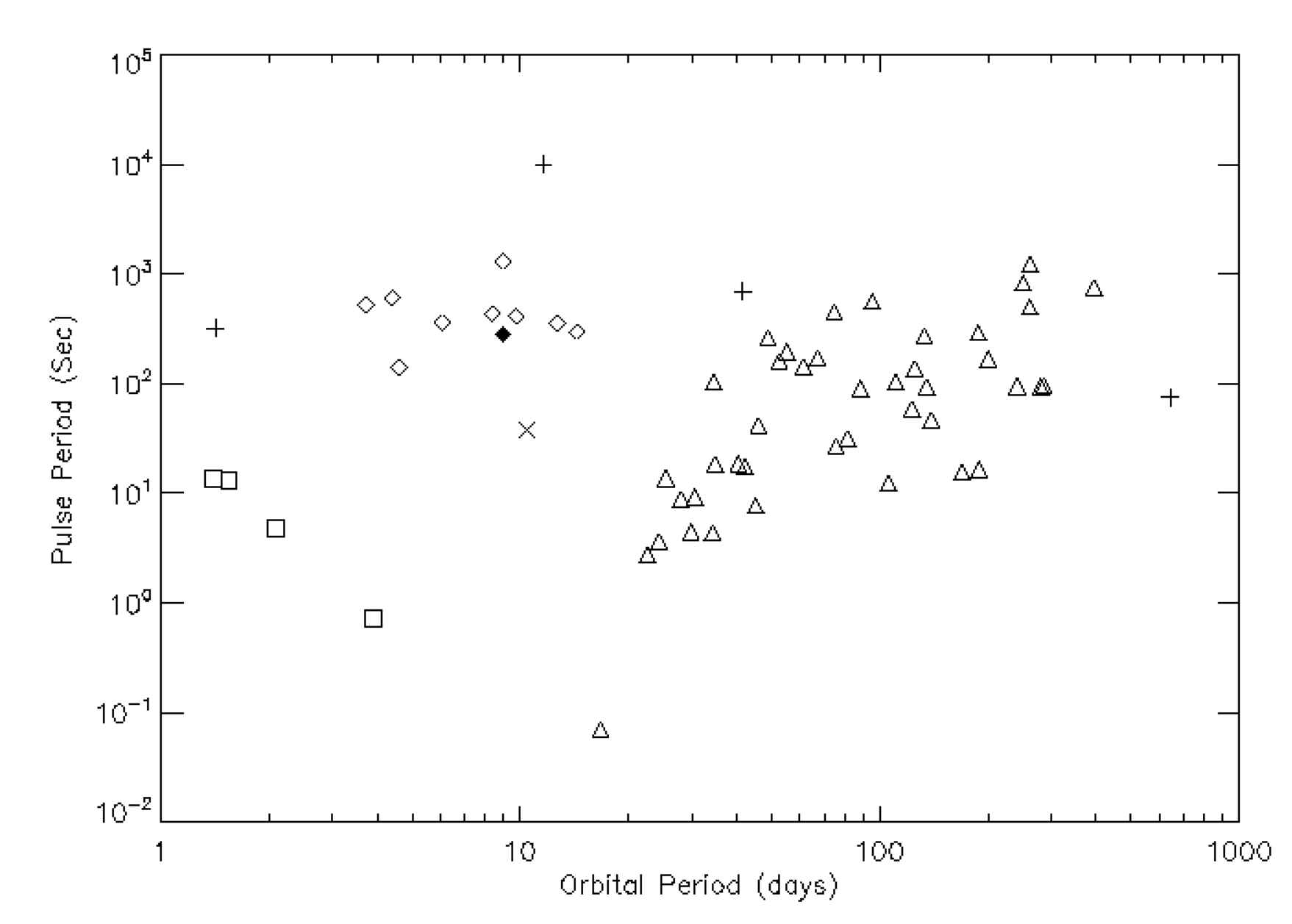}
\caption{Corbet diagram marking position of OAO 1657-415 and other HMXBs. OAO 1657-415 is marked by an X, EXO 1722-363 by a filled diamond. SGXRB Roche-Lobe Overflow systems  (Squares), Be/X binaries (Triangles), SGXRB Wind-fed systems (Diamonds) and anomalous systems (+) }
\end{figure}

\section{Orbital solution for EXO 1722-363}

The orbital solution we have calculated was obtained from archival ESO VLT data. Using a small subset of the available archive data, (11 spectra taken at different epochs spanning a wide range of orbital phase, from a set of 104 in total) we were able to measure radial velocities and construct the orbital solution shown in Fig. 3 (left). These spectra were centered on 2.1$\mu$m, having an integration time of 700s, and were obtained using the SW MRes mode with a 0.6$^{\prime\prime}$ slit. This resulted in high S/N spectra at a resolution R $\approx$ 4200.    
The resulting NS mass that we have determined from our orbital solution for EXO 1722-363  is consistent with the canonical mass of 1.4~M$_{\odot}$ measured in most other eclipsing HMXBs, except for that in Vela X-1, (Quaintrell et al.,\ 2003). The NS mass range we have determined stems from a lower and upper limit obtained using the following constraints - {\it Lower} : the system is viewed edge on (i.e. {\it i} = 90$^\circ$), {\it Upper} : the donor star fills its Roche lobe. Utilising this orbital solution we find a NS mass range of 1.5 - 1.6 M$_{\odot}$ (Mason et al. 2010). In a similar way the measured mass and radius of the supergiant donor, $M \sim 
13 - 15$~M$_{\odot}$ and $R \sim 25 - 28$~R$_{\odot}$ is determined, and this lends support to the B0-1 Ia spectral classification that we  
previously found (Mason et al.,\ 2009).

\begin{figure}[h]
\centering
\includegraphics[width=14cm]{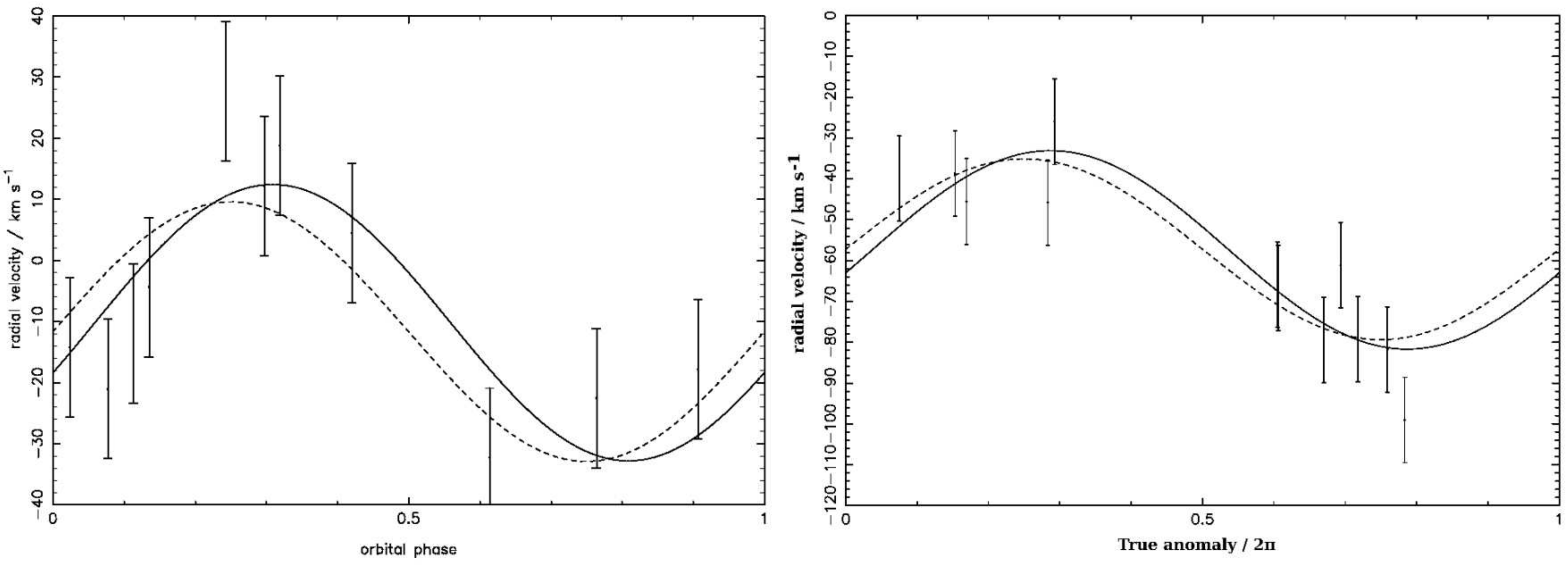}
\caption{Left : Radial velocity data for the donor in EXO~1722--363. Right : Radial velocity data for the donor in OAO 1657-415. In both cases the solid line is the best fitting sinusoid with three free parameters, the dashed line is that with a fixed zero phase in line with the published ephemeris. In the case of EXO~1722--363 the orbital phase is based upon the ephemeris of Thompson et al, 2007. For OAO 1657-415 the orbital phase is based upon the ephemeris of Bildsten et al, 1997. }
\end{figure}

\section{Orbital solution for OAO 1657-415}

As the mass donor in OAO 1657-415 is faint (H $\sim$ 11.7) we employed the NIR spectrometer ISAAC on the VLT to obtain high resolution (R $\sim$ 3000) and S/N spectra in the H band. 
Observations were conducted between 2008 May 13th and 2008 September 25th in the SW MRes  mode with a 0.8$^{\prime\prime}$ slit. Cross-correlation was performed using the standard IRAF routine {\it fxcor}. 12 high quality spectra were obtained that covered a wide range of orbital phase, sufficient to determine a dynamical mass solution for OAO 1657-415 (Fig. 3).    
Utilising this orbital solution we find a NS mass range of $\approx$  1.4 - 1.7 M$_{\odot}$  with a corresponding mass range for the counterpart star of $\approx$ 14 - 17 M$_{\odot}$. For a more precise mass determination please refer to Mason et al,\ 2011.


\section*{Acknowledgements}
ABM acknowledges support from an STFC studentship. JSC acknowledges support from an RCUK fellowship. 
This research is partially supported by grants AYA2008-06166-C03-03 and
Consolider-GTC CSD-2006-00070 from the Spanish Ministerio de Ciencia e
Innovaci\'on (MICINN).Based on observations carried out at the European Southern Observatory, Chile through programmes 081.D-0073(A and B) and 077.B-0872(A) .

%
%
\footnotesize
\beginrefer

\refer Audley, M.~D., Nagase, F., Mitsuda, K., et al., 2006, MNRAS, 367, 1147


\refer Bibby, J.~L., Crowther, P.~A., Furness, J.~P., et al., 2008, MNRAS, 386, 23

\refer Bildsten, L., Chakrabarty, D., Chiu, J. et al. 1997, ApJS, 113, 367

\refer Chakrabarty, D., Grunsfeld, J.~M., Prince, T.~A., et al, 1993, ApJ, 403, L33 Martins, F.

\refer Chakrabarty, D., Wang, Z., Juett, A.~M., et al., 2002, ApJ, 573, 789

\refer Corbet, R.~H.~D., Thorstensen, J.~R., Charles, P.~A. et al, 1986, MNRAS, 220, 1047

\refer Corbet, Robin H. D., Markwardt, Craig B., Swank, Jean H., 2005, ApJ, 633, 377. 

\refer Crowther, P.~A., Lennon, D.~J., Walborn, N.~R., 2006, A\&A, 446, 279

\refer Hanson, M.~M., Conti, P.~S., Rieke, M.~J, 1996, ApJS, 107, 281

\refer Hanson, M. M., Kudritzki, R.P., Kenworthy, M. A., et al, 2005, ApJS, 161, 154

\refer Martins, F., Genzel, R., Hillier, D.~J. et al, 2007, A\&A, 468, 233 

\refer Mason, A.~B., Clark, J.~S., Norton, A.~J., et al., 2009, A\&A, 505, 281

\refer Mason, A.~B., Norton, A.~J., Clark, J.~S. et al, 2010, A\&A, 509, 79

\refer Mason, A.B., Norton, A.~J., Clark, J.~S. et al, 2011, Submitted. 

\refer Meynet, G., Maeder, A., 2000, A\&A, 361, 101

\refer Morris, P.~W., Eenens, P.~R.~J., Hanson, M.~M. et al, 1996, 470, 597

\refer Quaintrell, H., Norton, A.~J., Ash, T.~D.~C. et al, 2003, A\&A, 401, 313 

\refer Thompson, Thomas W. J., Tomsick, John A., in 't Zand, J. J. M., et al., 2007, ApJ, 661, 447. 

\refer Warwick, R. S., Norton, A. J., Turner, et al., 1988, MNRAS, 232, 551

\refer Waters, L.~B.~F.~M., van Kerkwijk, M.~H. et al, 1989, A\&A, 223, 196
    

\refer Zurita Heras, J. A., de Cesare, G., Walter, R., et al., 2006, A\&A, 448, 261

\endrefer

\end{document}